\documentclass[twocolumn,twoside,slac]{revtex4}
\usepackage{graphicx}
\usepackage{fancyhdr}
\begin{document}

\title{Configuration of the ATLAS Trigger System}

%

\author{M. Elsing \footnote{Presenter at the conference}}
\affiliation{CERN, Division EP, 1211 Geneva 23, Switzerland}
\author{T. Sch\"orner-Sadenius}
\affiliation{CERN, Division EP, 1211 Geneva 23, Switzerland and\\Hamburg University, Institut f\"ur Experimentalphysik, Luruper Chaussee 149, 22761 Hamburg, Germany}
\author{\vspace{3mm} On behalf of the Atlas High Level Trigger Group~\cite{gaga}}

\begin{abstract}
In this paper a conceptual overview is given of the software foreseen to configure
the ATLAS trigger system. Two functional software prototypes
have been developed to configure the ATLAS Level-1 emulation and the
High-Level Trigger software. Emphasis has been put so far
on following a consistent approach between the two trigger systems
and on addressing their
requirements, taking into account the specific use-case of the
`Region-of-Interest' mechanism for the ATLAS Level-2 trigger. In the future
the configuration of the two systems will be combined to ensure a
consistent selection configuration for the entire ATLAS trigger system.
\end{abstract}

\maketitle

\thispagestyle{fancy}



\section{INTRODUCTION}

\begin{figure}
\includegraphics[width=82mm]{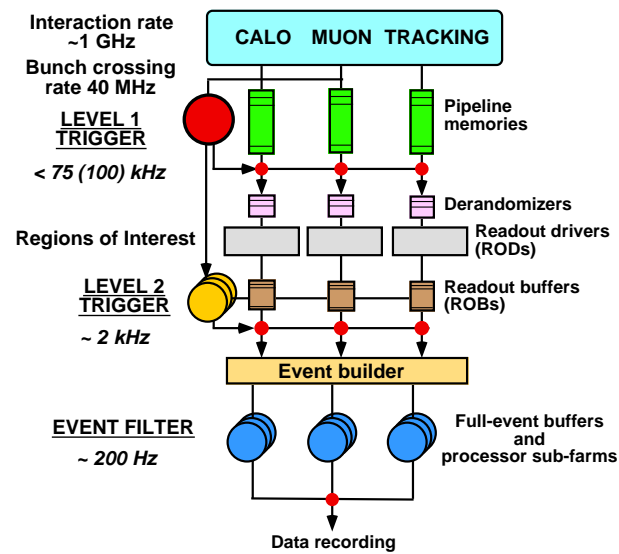}
\caption{\label{atlastrigger}A schematic view of the ATLAS trigger system.}
\end{figure}

The Large Hadron Collider (LHC), which is currently being built at the 
European Organization for Nuclear Research (CERN) in Geneva, 
will collide proton 
beams at a centre-of-mass energy of 14~TeV and with
a bunch-crossing rate of 
nominally 40~MHz. At the design luminosity 
of ${\rm 10^{34}~cm^{-2}s^{-1}}$ an average of about
25 proton proton interactions will take place in each of the 
bunch-crossings. An efficient and selective  trigger system is needed
to reduce the amount of data that will arise from these conditions
and to select the relevant physics events from the background of
soft interactions.  The trigger of the  ATLAS experiment~\cite{atlas}
is designed as a multi level system that reduces the event rate from 40~MHz
to about 200~Hz at which  events (that will have an average
size of about 1.6~MB)
can be written to mass  storage. Fig.~\ref{atlastrigger} gives an overview of
the trigger system which is divided in three levels (from top to bottom in 
Fig.~\ref{atlastrigger}) : 

\begin{itemize}
\item
The Level-1 (LVL1) trigger is a
hardware-based system that receives signals from the calorimeter and muon 
detectors of ATLAS. Its task is to reduce the event rate to 75~kHz within a 
latency of 2.5~$\mu$s. During that time the data from all detectors
are stored in pipelined memories. LVL1-accepted events are transfered to the 
Read-Out Buffers.
\item
The Level-2 (LVL2) trigger, which forms part of the High-Level 
Trigger (HLT), is based on software selection algorithms running in
processor farms. LVL2 can access data from all sub-detectors of ATLAS
in so called `Regions-of-Interest' that were identified by the LVL1 system.
The average time budget of LVL2 is about 10~ms, hence a fast
rejection strategy is needed using specialized trigger algorithms.
\item
The Event Filter (EF) is also based on software selection algorithms. In
contrast to LVL2 it runs after the event building, such that the
complete event information is available to the EF algorithms.  In the EF, a
thorough event selection and classification process will be
performed within a time
budget of a few seconds. The EF algorithms are foreseen to be based on offline reconstruction code using
the full calibration and alignment information. Events accepted by the EF
are written to mass storage.
\end{itemize}

In this paper, the concepts and mechanisms that are foreseen
for configuring the ATLAS trigger selection are discussed. The task
comprises the definition 
of so called `trigger menus' (i.e., the definitions of the
physics signatures the experiment
should be triggered on), the set-up of the selection software for the HLT as well as 
the set-up of the LVL1 trigger hardware. Currently
the latter part of the configuration is implemented only in a functional prototype
of the LVL1 emulation software.


\section{\label{lvl1}LVL1 TRIGGER CONFIGURATION}

\subsection{A short Overview of the LVL1 Trigger}

\begin{figure}[t]
\includegraphics[width=74mm,angle=-90]{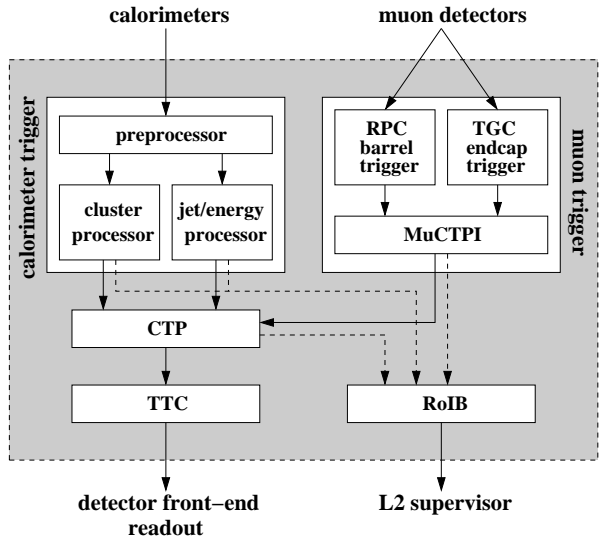}
\caption{\label{lvl1trigger}An overview of the LVL1 trigger system.}
\end{figure}

The LVL1 trigger~\cite{lvl1cite} is a complex hardware system consisting of
a calorimeter trigger, a muon trigger and the central trigger 
processor (CTP).  An overview of the LVL1 system\footnote{Note that the
Region-of-Interest Builder (RoIB) is formally not a LVL1 sub-system.}  is given in
Fig.~\ref{lvl1trigger}.

The calorimeter trigger receives as inputs
\~7200 analogue signals from a dedicated trigger-tower electronics that
locally combines information from calorimeter cells in the various
ATLAS calorimeters. A trigger tower has a typical granularity of
$\Delta\eta \times \Delta\phi = 0.1 \times 0.1$. 

The task of the calorimeter trigger is to search for localized energy
depositions that are a signals for high transverse energy
electrons/photons, $\tau$ particles,
hadrons or jets. The energy depositions are compared 
to a set of programmed transverse-energy 
thresholds and the multiplicity of objects passing each of the thresholds
is counted. There are $16-N$ thresholds for electrons/photons, $N$ for $\tau$
leptons or hadrons, eight for central jets, and four for forward jets.
In addition, the calorimeter 
trigger provides global energy sums to measure the total and missing
transverse energy in an event. These sums are discriminated against
eight thresholds for missing transverse energy and four for
total transverse energy.

The multiplicity for each threshold is sent to the CTP. 
In addition the type, position and threshold information about each
candidate object is recorded in so called `Regions-of-Interest' (RoIs).
For selected events, these RoIs are sent to the HLT via the
Region-of-Interest Builder (RoIB) in order to seed the LVL2 selection.

In analogy, the muon trigger, which is based on information from dedicated fast 
muon chambers, derives the multiplicity count for muon candidates passing 
six programmable thresholds. The multiplicities are again sent to the 
CTP, and the information on the candidates or RoIs is sent to the HLT. 

The CTP receives all the multiplicity and energy-sum-threshold
information from the calorimeter and 
muon triggers. It combines them according to the trigger menu to
derive the LVL1 event decision. The CTP provides information to the HLT for selected events indicating which signatures were fulfilled.


\subsection{Configuration of the LVL1 Trigger}

For the final system the task of the LVL1 configuration will be the 
preparation of the LVL1 hardware for data taking. Up to now only preliminary
ideas exist on how to configure consistently the LVL1 sub-system
hardware, the calorimeter and muon triggers and the CTP.
Questions being addressed in this context range from  storage of hardware
configuration files, through keeping track of configuration changes,
to checks for consistency and hardware compliance. These issues will not be
touched upon in this article. However, the configuration of the LVL1 has a purely
algorithmic aspect that defines the physics selection strategy. This part of the
configuration, which is implemented in the LVL1 software prototype and is
part of the ATLAS simulation chain, is described here.

The first task of the LVL1 configuration code is to translate the trigger menu,
 i.e. the collection of physics signatures LVL1 is supposed to trigger on,  into
something that the emulation of the CTP can understand and use in making
the event decision based on the calorimeter and muon trigger inputs.
A physics signature, a so called `trigger item', is a logical combination of 
requirements formulated as `trigger conditions' in terms of multiplicities of
candidate objects delivered by the calorimeter and muon triggers.
Such candidates are typically high-$p_T$ objects in the events that
are a signature for interesting physics events.
A simple example of a trigger item is:

\begin{center}
\begin{em}
$\ge$ electron/photon candidate with $E_T >$~10 ~GeV \\ and
\\ $\ge$ muon candidate with $p_T >$~15~GeV
\end{em}
\end{center}

\noindent
or, in the ATLAS LVL1 notation:

\begin{center}
1EM10+1MU15 
\end{center}

\noindent
which combines two trigger conditions `1EM10' and `1MU15'. In this notation
the string `EM' represents a candidate of type electron/photon and the
integer numbers before and after the string symbolize the required multiplicity
and transverse energy/momentum, respectively. The combination of a candidate
string and a threshold value like `EM10' is called a `trigger threshold'. The ATLAS
LVL1 system distinguishes electron/photon (`EM'), muon (`MU'), tau/hadrons (`HA'),
(forward) jets (`JT', `FR', `FL'), total and missing transverse energy (`ET', `TM'), and
the transverse  energy calculated as the sum of all jet transverse energies (`SM').

The second task of the configuration code is to set up the the calorimeter and muon
trigger sub-systems such that they deliver the information required by the CTP
in order to derive the event decision based
on the trigger menu. The LVL1 sub-detector simulation has to be configured so
that each sub-system delivers the multiplicities for the correct set of
trigger thresholds. For the above example, the
calorimeter trigger has to be configured such that it delivers
to the CTP the multiplicity count
for the threshold `EM10', i.e. the number of electron/photon candidates with
transverse energy above 10~GeV. It is obvious that  the trigger menu and the
trigger thresholds for the calorimeter and muon triggers have to be defined
consistently. In particular, all thresholds used in the definition of any trigger
condition in any trigger item must be delivered by the calorimeter and muon
trigger simulations and thus need to be configured.

In the configuration process for the CTP simulation the restrictions imposed by
limited abilities and resources of the real hardware have to be taken into account.

The LVL1 trigger configuration software is currently being adapted to configure
as well the LVL1 trigger hardware by deriving the necessary look-up table and
FPGA configuration files from the trigger menu and trigger threshold list. Such 
a common configuration scheme will allow for cross-checks between hardware
and software.


\subsection{\label{thresholds}XML Definition of Trigger Thresholds}

To configure LVL1 the trigger menu and the list of thresholds are defined 
using XML and are parsed into instances of C++ classes using the Xerces DOM API~\cite{xerces}. The chosen notation exploits the facility of XML to define logical
structures by introducing  user-defined tags. The tag structure used for the `trigger
thresholds' is:

\begin{footnotesize}
\begin{verbatim}
<TriggerThresholdList>
   <TriggerThreshold name="..." type="..."> 
      <TriggerThresholdValue thresholdval="..." /> 
   </TriggerThreshold>
</TriggerThresholdList>
\end{verbatim}
\end{footnotesize}

It is important to note that
a trigger threshold contains one or more `trigger threshold values'. 
This concept allows to assign different threshold values (in GeV) to various
topological regions of the detector using attributes that set limits to the validity
in terms of $\eta$ and $\phi$ in ATLAS. This concept is foreseen for the calorimeter
trigger hardware and is subject to current detailed studies. The previous
example using a set of trigger threshold values may look like:
 
\begin{footnotesize}
\begin{verbatim}
<TriggerThresholdList>
   <TriggerThreshold name="..." type="..."> 
      <TriggerThresholdValue 
          thresholdval="5" etamin="-2" etamax="0" /> 
      <TriggerThresholdValue 
          thresholdval="10" etamin="-5" etamax="-2" /> 
      <TriggerThresholdValue 
          thresholdval="5"  etamin="0" etamax="2" /> 
      <TriggerThresholdValue 
          thresholdval="10" etamin="2" etamax="5" /> 
   </TriggerThreshold>
</TriggerThresholdList>
\end{verbatim}
\end{footnotesize}

The $<$TriggerThreshold$>$ tag has several attributes, the most important ones 
are `name' and `type':

\begin{itemize}
\item The `name' attribute assigns a unique label to the trigger 
threshold. It is needed to connect the threshold to the trigger condition.
An example for a `name' is `EM10'.
\item The `type' attribute is required mainly for technical reasons: The number
of the different thresholds is limited - for example only up 6 muon thresholds
can be implemented because of hardware limitations. The `type'
attribute helps in this book-keeping. 
\end{itemize}

When the XML tags are parsed, the attributes are translated into data members 
of the corresponding C++ class instance.

When the XML tags are parsed, the attributes of the
$<$TriggerThresholdValue$>$ tags are translated into data members
of the corresponding C++ objects. Depending on the `type' attribute
of the $<$TriggerThreshold$>$ tag, different attributes for the
$<$TriggerThresholdValue$>$ tags are expected (e.g., definition of
isolation criteria). Only the `thresholdval'
attribute is common, which is used to define the threshold value (in GeV).


\subsection{\label{menu}XML Definition of the Trigger Menu}

The XML definition of the trigger menu uses the $<$TriggerMenu$>$, $<$TriggerItem$>$ and $<$TriggerCondition$>$ tags.
 The basic structure of the XML file for the trigger menu is the following:

\begin{footnotesize}
\begin{verbatim}
<TriggerMenu>
   <TriggerItem>
      <TriggerCondition threshold="..." mult="..."/>
   </TriggerItem>
</TriggerMenu>
\end{verbatim}
\end{footnotesize}

In addition the special tags $<$AND$>$, $<$OR$>$ and $<$NOT$>$ are available
to allow for logical combinations of trigger conditions for a given the
trigger item, for example:

\begin{footnotesize}
\begin{verbatim}
<TriggerMenu ...>
   <TriggerItem ...>
      <AND>
         <TriggerCondition ... />
         <OR>
            <TriggerCondition ... />
            <NOT>
               <TriggerCondition ... />
            </NOT>
         </OR>
      </AND>
   </TriggerItem>
</TriggerMenu>
\end{verbatim}
\end{footnotesize}

\begin{figure}
\includegraphics[width=75mm]{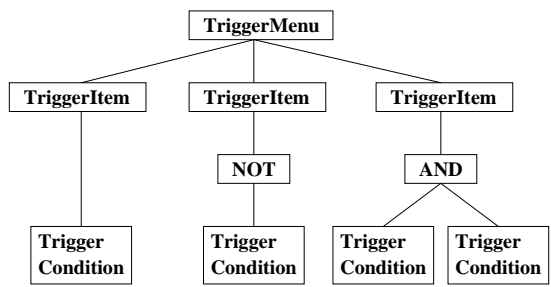}
\caption{\label{xmltree}Schematic overview of a simple XML tree build from a trigger menu file.}
\end{figure}

The $<$TriggerMenu$>$ tag has as only attribute `TM\_ID'  to set a label for the
trigger menu. The $<$TriggerItem$>$ has different attributes: `TI\_ID' is used
to set a name; `mask'  indicates via values `on' or `off' whether or not the
item is to be used in the LVL1 decision; `priority'  is set to `low' or `high',
depending on whether or not the item should have priority in the CTP
dead-time algorithm; `prescale'  is an integer indicating the prescale factor
to be used in the CTP simulation. The $<$TriggerCondition$>$ 
tag has two attributes: `threshold' to give it a label and a `mult' to specify the
required mulitplicity for this trigger condition. The `threshold' attribute has
to be the same as the `name' attribute of a corresponding
$<$TriggerThreshold$>$ tag. An example of the logical
structure of the XML tree of the trigger menu is shown in Fig.~\ref{xmltree}.

\subsection{Implementation of the LVL1 Configuration}

\begin{figure}[htb]
\includegraphics[width=65mm]{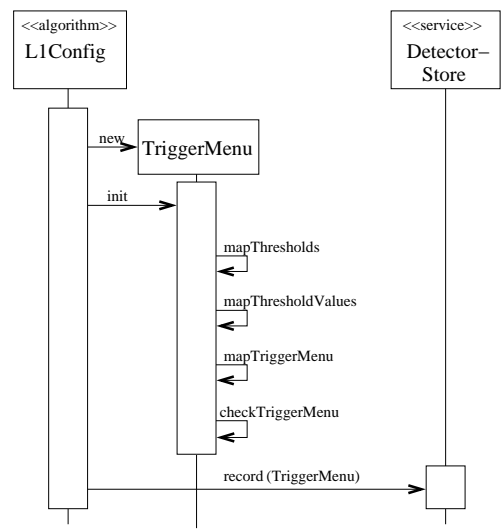}
\caption{\label{configseq}Sequence diagram for the main LVL1 configuration 
which sets up the {\em TriggerMenu} object. 
See text for details.}
\end{figure}

Fig.~\ref{configseq} shows a sequence diagram of the configuration process.
{\em L1Config} is the name of the algorithm in the ATLAS offline framework
Athena, in which the HLT trigger configuration is implemented.
The central class of the configuration is the so-called
{\em TriggerMenu}. A single 
instance is created by the {\em L1Config} algorithm, that also 
calls the {\em TriggerMenu::init} method.
The {\em TriggerMenu} 
contains or has access to all the information in the two XML files for the trigger 
menu and the trigger thresholds.

The LVL1 trigger menu basically is a collection of trigger items. Therefore
the {\em TriggerMenu} has as a data member a vector of
{\em TriggerItem} objects. In turn, each trigger item contains a vector
of {\em TriggerCondition} objects. The logical structure of the
item condition relations is not directly reflected in the C++ class
structure, but is available from the  XML tree that is created in memory
in the parsing process (see Fig.~\ref{xmltree}).

The {\em TriggerMenu::mapTriggerMenu} method is used to translate the 
trigger menu XML file into  {\em TriggerItem} and 
{\em TriggerCondition} objects. For each $<$TriggerItem$>$ or
$<$TriggerCondition$>$ tag, a {\em TriggerItem}
or {\em TriggerCondition} object is created and pushed into the corresponding
vectors.

\begin{figure*}[t]
\centering
\includegraphics[width=120mm]{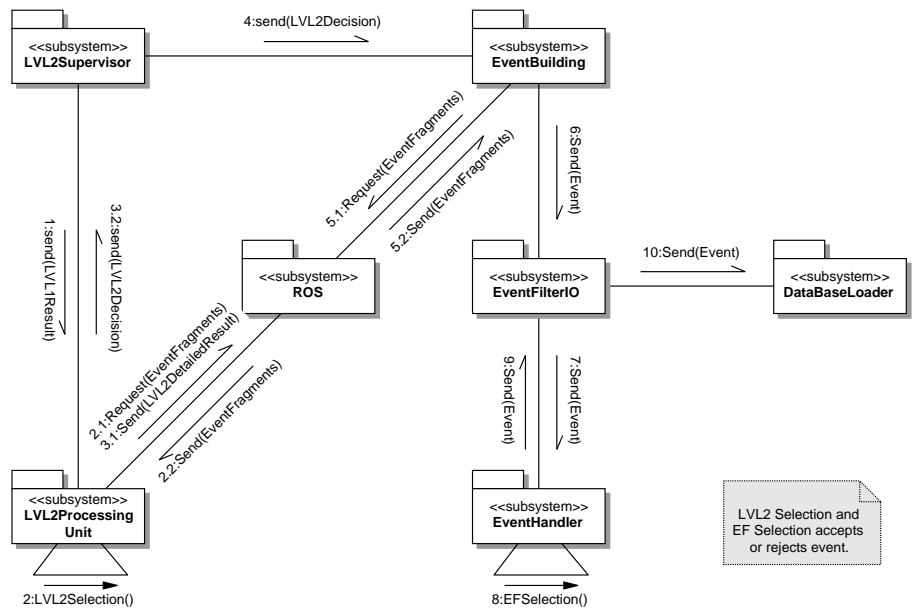}
\caption{\label{hltflow}An overview of the data flow in the HLT.}
\end{figure*}

The {\em TriggerMenu} holds a map to
combine the labels of the trigger thresholds with pointers to the 
{\em TriggerThreshold} objects that are created by the
{\em TriggerMenu::mapThresholds} method.
A {\em TriggerThreshold} in turn holds pointers to 
one or more {\em TriggerThresholdValue} objects that are
created in the  method {\em TriggerMenu::mapThresholdValues}.
The connection between {\em TriggerCondition} and {\em TriggerThreshold}
instances  is needed for the CTP simulation. It is implemented using string
comparisons between the data members that correspond to the
`name'  and `threshold' attributes of the 
$<$TriggerThreshold$>$  and the $<$TriggerCondition$>$ tags.

As a last step in the configuration, checks are performed on 
the {\em TriggerMenu} object in order to test its completeness and 
consistency (method {\em TriggerMenu::checkTriggerMenu}).
The {\em TriggerMenu} is afterwards recorded in the so-called `Detector Store'
that is provided by the offline framework.

The configuration process also covers configuration of the calorimeter and
muon trigger simulations. For this purpose {\em CTPCaloConfig},
{\em CTPJetEnergyConfig} and  {\em CTPMuonConfig}
objects are created that hold the thresholds to be delivered by the
different triggers.
	
The next development step will be to use the software to configure the existing
CTP demonstrator hardware. Once hardware and software are configured
from the same source, detailed tests
of the CTP hardware can be performed by comparing the simulated CTP result
to the hardware result for arbitrary test input patterns.


\section{\label{hlt}HLT CONFIGURATION}

\subsection{\label{hltoverview}HLT Trigger Overview}

Fig.~\ref{hltflow} shows an overview of the HLT data flow ~\cite{hlttp}. The
LVL2 Supervisor sends the LVL1 result containing `Region-of-Interest'
information (e.g. geometrical position of interesting objects identified by LVL1)
to a LVL2 Processing Unit. This unit
performs the LVL2 selection for the event. It retrieves event data in `Regions-of-Interest' from the Read-Out System (ROS). At the end of the selection process
the LVL2 decision is passed back to the Supervisor. In case of a positive trigger
decision, the LVL2 detailed result is sent to a dedicated LVL2 ROS. The
LVL2 Supervisor sends the decision to the Event Builder, which assembles
the full event. The complete event is sent to the Event Filter IO and from there
to one of the Event Handlers, which performs the EF selection. EF-selected
events are written to mass storage.

\begin{figure}[b]
\centering
\includegraphics[width=85mm]{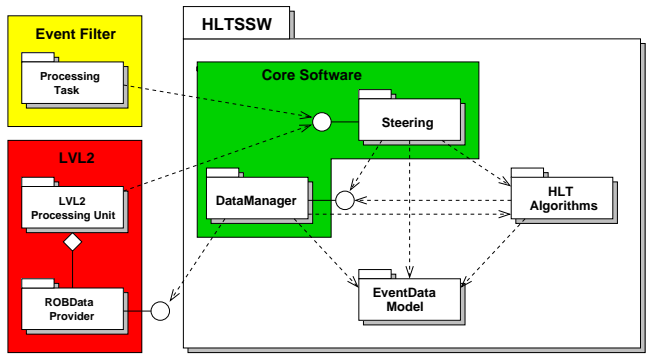}
\caption{\label{hltssw}A package view of the HLT selection software.}
\end{figure}

Fig.~\ref{hltssw} shows a package view of the Event Selection Software~\cite{elsing}
which runs in the LVL2 Processing Unit and in the EF Event Handler. It has four
building blocks: the Event Data Model defines the structure of the event;
the Data Manager handles the event data and is used in LVL2 to retrieve
the required raw data on demand from the ROS; the HLT Algorithms provide
the algorithmic part of the selection; the Steering controls the full selection
process. It is the task of the HLT configuration to configure the Steering.

\subsection{\label{hltprinciple}The Principle of the HLT Selection}

The HLT selection software provides a common framework to implement
the LVL2 and EF selection. The software is based on two concepts that
are designed to minimize the latency and the amount of raw data to be
moved and analyzed. LVL1 provides for each event
a set of `Region-of-Interest' coordinates
via the RoI Builder. These RoIs are used by the
LVL2 Steering to seed to algorithmic processing and to restrict the
data access to the regions in
the detector identified by the LVL1 system.

The 2nd concept is the so-called
`step' processing. The sequence of HLT algorithms to verify a given LVL1 RoI
is sub-divided into several logical steps. This allows the Steering to execute
in the first step for each RoI those algorithms giving a large rejection power
for little data movement and processing.
In each subsequent step it is the
task of the HLT algorithms to refine the event information using an increasing
amount of additional data from different ATLAS sub-detectors. It is important
to note that the Steering will execute the first step for all RoIs in one go.
Events may be rejected after each step; remaining events are passed to
the next step for further processing.
For events accepted after the final step a
positive trigger decision is transmitted to the LVL2 Supervisior.

The implementation of the Steering of the HLT selection software must to
be generic. The HLT selection is data driven by the LVL1 RoIs, while the
list of possible HLT algorithm sequences will be configured at run time based on
the criteria of the physics signatures in the trigger menu. It was found to be
beneficial to abstract the selection process in terms of so-called `Trigger Elements'
(TEs). A TE represents a trigger relevant object in the event, for example
a LVL1 RoI. In this picture, the HLT selection can be seen as a refinement
process of TEs. Each TE implies by its label a physics interpretation, like for example
an `isolated electron with $p_T > 20$~GeV'  or `e20i'. Each physics signature
in a trigger menu is defined as a combination of required TEs,
for example `e20i + mu30i'.

The seeding of the HLT Algorithms is implemented by navigation from the TE
to the relevant event data (RoI, clusters, tracks, ...). Thereby the Steering only
analyses the TE content of an event and requests algorithm processing to
refine the information. The details of the specific event data related to each
LVL1 RoI are only visible to the concrete HLT algorithm.


\subsection{Signatures and Sequence Tables}

The basic unit of a step is a `sequence'. It consists of a list of input TEs,
a list of HLT algorithms to be run on each set of matching input TE combinations
found in an event, and one (and only one) output TE that represents
the hypothesis of the expected reconstruction result.
The complete set of sequences to be run in a given step is called the
`sequence table'.

At the end of each step the Steering compares the list of validated
TEs in an event to so-called `signatures', which are formulated as
required TEs or TE combinations. The set of signatures of a given
step is called the `menu table'. The signatures in the menu table of the
final step are called `physics signatures'.
The physics signatures are defined according to the ATLAS physics
program and are the ones that will be visible to
the shift crew in the control room, whereas the intermediate ones are
visible only for an expert with insight into the configuration
scheme.

From the above it is clear that the trigger configuration has to
provide a set of menu and sequence tables. This is achieved in a
top-down approach starting from a list of physics
signatures that is specified for a given ATLAS run. This list is used
to derive, in a recursive way, all necessary sequence and menu tables.

\begin{figure}
\includegraphics[width=70mm]{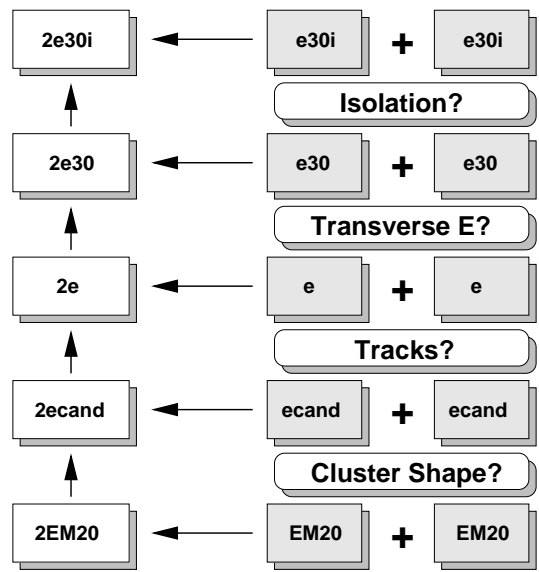}
\caption{\label{hltconfig}A configuration example. See text for details.}
\end{figure}

The recursive procedure is illustrated for a simplified example in
Fig.~\ref{hltconfig}. The physics signature is `two isolated 
electron candidates with $E_T >$~30~GeV' or  `2 e30i'.
This signature requires the presence of two final TEs `e30i' in an event.
The configuration
software has a list of all implemented sequence and checks which sequence
has this TE as its output. From the matching sequence is determines the
corresponding required input TEs. In the example the `e30i' TE is
made from an `e30' TE using an isolation algorithm. Thereby the
configuration can derive that two TEs `e30' are required
in the step that leads to the physics signature `2e30i'. The resulting 
intermediate signature is `2e30'. Similarly, the `2e30' is composed of two
`e30' TEs  which are made from TEs labeled `e' in an algorithm that
requires a minimum transverse energy. In this manner, all intermediate signatures  and also the necessary sequences are derived recursively, until the configuration 
algorithm arrives at input TEs labeled `EM20' that represent the LVL1
electromagnetic RoI objects.

It is worth mentioning that the HLT event decision is derived from the same
logical structure, but now using it bottom-up: The HLT selection process starts
from the input from LVL1. In the example it would search for two
`EM20' TEs. In case these TEs are not present, the signature of the first step (`2EM20')  cannot be fulfilled and the event would be rejected. If the signature `2EM20' is fulfilled, the sequence of algorithms is executed on each input TE to analyze the cluster shape of the candidate object.
The task is to refine the electron selection, rejecting background (mainly jets)
using clusters shape variables.
In other words, for each `EM20' TE that is
considered to be due to an electron, the hypothesis of a TE `ecand' (for `electron candidate') needs to be validated. If an `ecand' is validated for 
both `EM20' TEs, the signature `2ecand' is fulfilled and the selection will proceed.
Otherwise the event will be rejected. In this way the selection is carried out up
to the last step that ends with two TEs `e30i'.


\subsection{\label{hltimplementation}Implementation of the HLT Configuration}

\begin{figure}
\centering
\includegraphics[width=75mm]{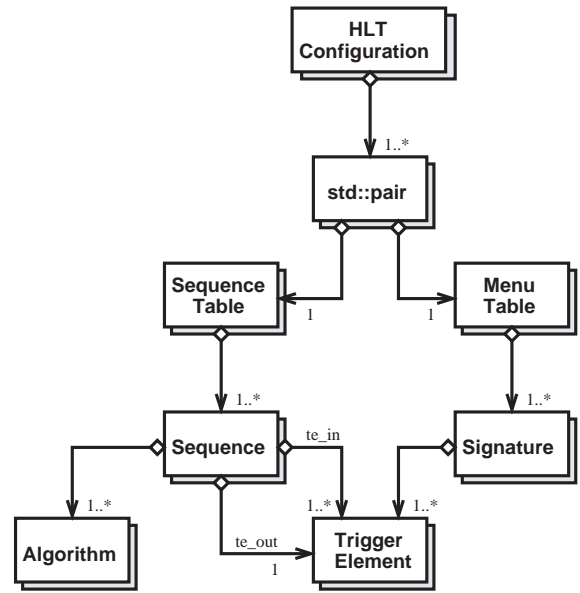}
\caption{\label{hltuml}A UML class diagram of the HLT configuration software.} 
\end{figure}

As in the case of the LVL1 trigger, two XML files are used to define the
physics signatures and the list of available sequences for the HLT
configuration. The signature file contains
$<$SIGNATURE$>$ tags that have the a list of $<$TRIGGERELEMENT$>$
tags for the required TEs:

\begin{footnotesize}
\begin{verbatim}
<SIGNATURE>
   <TRIGGERELEMENT te_name="..." />
   <TRIGGERELEMENT te_name="..." />
</SIGNATURE>
\end{verbatim}
\end{footnotesize}

The `te\_name' attribute assigns a label to the TE. The physics menu in
the XML file is given by the list of all signatures. The second XML file
contains a list
of $<$SEQUENCE$>$ tags. Each sequence has three different attributes:

\begin{footnotesize}
\begin{verbatim}
<SEQUENCE te_in="..." algo="..." te_out="..." />
\end{verbatim}
\end{footnotesize}

\noindent
to identify the input TEs (`te\_in'), 
the algorithm(s) to be run on these trigger elements (`algo'), and the output
TE (`te\_out'). The same `te\_in' can appear in more than one
sequence.

The XERCES DOM API~\cite{xerces} is used to parse
the $<$SIGNATURE$>$ and $<$SEQUENCE$>$ tags
in the configuration XML files into an object tree. Afterwards the
recursive algorithm described above starts to derive the full set of
menu and sequence tables. In practice
the menu and sequence tables need additional processing to allow for
a realistic menu that consists of several physics signatures. Furthermore,
the borderline between LVL2 and EF is defined by assigning the
sequences of the first set of steps to LVL2 and the remaining ones
to the EF.

Fig.~\ref{hltuml} shows a UML class diagram of the HLT Configuration used 
to store the HLT configuration information. It is based on a collection of
pairs, one instance per step. Each pair combines a {\em MenuTable}
and a {\em SequenceTable} object, each of  which holds a vector
of {\em Sequence} or {\em Siganture} objects, respectively.
A {\em Signature} combines several {\em TriggerElements} and
a {\em Sequence} holds one output {\em TriggerElement} , one
or more  input {\em TriggerElements} and a list of {\em Algorithms}.

The HLT configuration has been tested as part of the HLT selection software
in the offline and in dedicated online test-beds. In the future it is foreseen
to combine the configuration of the HLT and LVL1, based on
the input menu table of the first HLT step. Thereby a consistent
configuration of the complete ATLAS trigger system will be achieved.


\section{CONCLUSION}

An overview of the ATLAS trigger system has been given, with emphasis on 
the configuration of the various trigger levels, namely the LVL1 trigger and
the HLT. For LVL1 efforts so far concentrated
on the configuration of the trigger simulation, with the aim of a common
configuration  for both the hardware and the simulation software. In the case
of the HLT, which is implemented as a software trigger running
in processor farms, the configuration
software has been tested successfully and used offline and in dedicated
online test-beds. Work is now in progress to combine the configuration of the LVL1 trigger and the HLT in order to guarantee a consistent and efficient ATLAS trigger
selection strategy.

\begin{acknowledgments}
We would like to thank the ATLAS LVL1 Trigger and Data Acquisition Groups for
their contributions to this work and their help in preparing this paper.
\end{acknowledgments}



\begin{footnotesize}
$^{*1}$
S. Armstrong, J.T. Baines, C.P. Bee, M. Biglietti, A. Bogaerts, V. Boisvert, M. Bosman, S. Brandt, B. Caron, P. Casado, G. Cataldi, D. Cavalli, M. Cervetto, G. Comune, 
A. Corso-Radu, A. Di Mattia, M. Diaz Gomez, A. dos Anjos, J. Drohan, N. Ellis, 
M. Elsing, B. Epp, F. Etienne, S. Falciano, A. Farilla S. George, V. Ghete, S. Gonz‡lez, M. Grothe, A. Kaczmarska, K. Karr, A. Khomich, N. Konstantinidis, W. Krasny, W. Li, A. Lowe, L. Luminari, H. Ma, C. Meessen, A.G. Mello, G. Merino, P. Morettini, 
E. Moyse, A. Nairz, A. Negri, N. Nikitin, A. Nisati, C. Padilla, F. Parodi, V. Perez-Reale, J.L. Pinfold, P. Pinto, G. Polesello, Z. Qian, S. Rajagopalan, S. Resconi, S. Rosati, 
D.A. Scannicchio, C. Schiavi, T. Schoerner-Sadenius, E. Segura, T. Shears, 
S. Sivoklokov, M. Smizanska, R. Soluk, C. Stanescu, S. Tapprogge, F. Touchard, 
V. Vercesi, A. Watson, T. Wengler, P. Werner, S. Wheeler, F.J. Wickens, 
W. Wiedenmann, M. Wielers, H. Zobernig
\end{footnotesize}

\end{document}